\documentstyle[12pt]{article}

\font\sqi=cmssq8
\def\DR{\rm I\kern-1.45pt\rm R}
\def\DC{\kern2pt {\hbox{\sqi I}}\kern-4.2pt\rm C}
\renewcommand{\thefootnote}{\fnsymbol{footnote}}
\begin{document}
\thispagestyle{empty}
{\hfill  Preprint JINR E2-92-411}\vspace{2.5cm} \\
\begin{center}
{\large\bf Even and Odd Symplectic and K\"ahlerian Structures
          on  Projective Superspaces } \vspace{1.5cm}
\\
{\large O. M. Khudaverdian} \vspace{0.5cm} \\
{\it Yerevan State University  } \\
{\it Yerevan, Armenia } \vspace{1cm}\\
{\large A. P.
Nersessian\footnote{E-MAIL:NERSESS@THEOR.JINRC.DUBNA.SU}}\vspace{0.5cm}\\
{\it Laboratory of Theoretical Physics,} \\
{\it Joint Institute for Nuclear Recearch}\\
{\it Dubna, Head Post Office, P.O.Box 79, 101 000 Moscow, Russia}
\vspace{2cm} \\
{\bf Abstract}
 \end{center}

Supergeneralization of $\DC P(N)$ provided by even and odd K\"ahlerian
structures  from Hamiltonian  reduction are construct.Operator $ \Delta$ which
used in Batalin-- Vilkovisky quantization formalism  and mechanics which are
bi-Hamiltonian under corresponding even and
odd Poisson brackets  are considered.

\begin{center}
{\it Submitted to J. Math. Phys.}
\end{center}

\vfill
\setcounter{page}0
\renewcommand{\thefootnote}{\arabic{footnote}}
\setcounter{footnote}0
\newpage
   \setcounter{equation}0
	    \section{Introduction }
  On the supermanifolds it is possible to define not only even, but also odd
 symplectic structures [1].
    Phase space structure corresponded
  to even symplectic structure and odd one.

  For example on the superspace  $E^{2N,M}$   with
    coordinates $z^A = (x^1,...,x^{2N}, \theta^1,..., \theta^{M})$
    one can consider an even symplectic structure  with corresponding
  even canonical Poisson bracket:
   \begin{equation}
		  \{f,g\}_0
		     =
			\sum_{i=1}^N \left(
	       \frac{\partial f}{\partial  x^i}
	       \frac{\partial g}{\partial x^{i+N} }
		       -
		\frac{\partial f}{\partial  x^{i+N}}
		\frac{\partial g}{\partial  x^i}
			   \right)
			   +
			  \sum_{\alpha =1}^{M}
	    \epsilon _\alpha \frac{\partial^{R} f}{\partial  \theta^ \alpha}
		\frac{\partial^{L} g }{\partial  \theta^{\alpha}} ,\;\;\epsilon_{\alpha} =\pm
1 .
   \end{equation}
On $E^{N.N}$ one can consider an odd one with corresponding canonical
   odd Poisson bracket (Buttin bracket,
   antibracket):
   \begin{equation}
		  \{f,g\}_1 =
			\sum_{i=1}^{N}\left(
		\frac{\partial f}{\partial  x^i}
		\frac{\partial^{L} g}{\partial\theta^i}
			 +
              \frac{\partial^{R} f}{\partial  \theta^i}
		\frac{\partial g}{\partial  x^i}
		   \right)
\end{equation}
  In the [2, 3] Batalin and Vilkovisky used
    odd bracket  for formulating  Lagranian BRST
   quantization formalism (BV -formalizm). Its provides a possibility to give
covariant and the most elegant formulation of the
     conditions on all the ghosts . BV-formalism is an effective
method
for quantization of gauge theories with open Lie algebra. An attempt to
consider it as a framework of background independent open-string field theory
was made [4, 5].

On other hand  the possiblity importance  of the odd bracket
      in twistorial program  and supersymmetric mechanics
     was emphasised [6-9].
     The problem of reformulation of  supersymmetric mechanics in
     terms of odd bracket, using the supercharge as a new
     Hamiltonian and the attempts to quantize it were performed in [8-10].

    There is no doubt that  odd bracket   needs to be  geometrically
    investigated.

It is possible to  formulate Hamiltonian mechanics  in term of odd
brackets  as well as in term of even one [11]. Arbitrary even
nondegenerate bracket  can be reduced (locally) to canonical form (1.1), and
arbitrary odd one -- to canonical form (1.2) [12]. But in the general case even
and odd brackets cannot be simultaneously reduced to form (1.1) and (1.2).
   The structure of the supergroup of
   transformations which preserve both brackets, depends on their
   mutual position. Anyway this supergroup is finite-dimensional
  and it is the different grading of the brackets that leads to
  this fact [13].

There are nontrivial geometrical
  objects depending on second derivatives which are invariant
  under transformations preserving odd bracket and the volume
  form connected with even bracket. It is the "operator $\Delta $" [13]
 which used in BV -formalism [2, 3]
  and the semidensity constructed in [14].
   These objects have no analogs in a classical case.

   These results strongly indicate that nontrivial
   geometry arises on the supermanifolds which are provided by
   Poisson brackets of different grading. Geometrical properties  of
   superspaces provided	by Poisson brackets of different gradings were
    investigated in [13, 15-17]. Superspaces, provided simultaneously by even
 and odd  canonical one was investigated in [15].
  It was shown in [16] that exists a large class of supermanifolds
  (the supermanifolds, assotiated with tangent bundles of K\"ahlerian
  manifolds) on which one can defined simultaneously even add odd
  symplectic
 (and K\"ahlerian ) structures. These
   structures turn out to be lifting  of the corresponding
   structures on the underlying manifolds. Therefore their properties
    have to be expressible in terms of classical geometrical
  objects.  They are good models for revealing geometrical  properties of
   two -bracket supermanifolds.

But there don't support by nontrivial  examples where even and odd
symplectic structures have natural geometrical origin.\\

In this work  we
construct the example of  such   supermanifold as
   reduced phase superspaces of the superoscillator.
(The dynamics
   of the superoscillator in the superspace $E^{2N,2N}$ can be described
   either in terms of the canonical even bracket (1.1) or in terms of the
   canonical odd one (1.2). In the second case the role of the Hamiltonian
   is played by one of its supercharges [15].)

    In the {\it Section 2} we demonstrate reduction procedure
   on the simple examples constructing phase superspaces
   reduced by the Hamiltonian of the superoscillator.
    This procedure performed in terms of even and odd structures
  leeds to the two different supermanifolds .

      In the {\it Section 3} we perform
     the reduction procedure by Hamiltonian of the superoscillator
   and by its supercharges.
   In terms of both symplectic structures we come to
   the same supermanifold which naturally inherits even and odd
   structures of the initial superspace. Canonical complex
   structure on the initial superspace $E^{2N,2N}$
   provides this supermanifold with the complex structure and
   with the even and odd K\"ahlerian structures corresponding to them.

     It  occurs that this supermanifold is associated to the
   tangent space of the underlying manifold - complex projective space.

    This supermanifold obtained by reduction procedure
  can be naturally included in the family of the supermanifolds
( which are associated with tangent bundles  of arbitraty K\"ahlerian
 manifold) with even and odd K\"ahlerian structures lifted
 from the K\"ahlerian structure of the underlying manifold, which was
investigated in [16].

 In the {\it Section 4} we investigate the bi-Hamiltonian mechanics
 (i.e. the  even vector fields preserving both symplectic
 structures)  and "operator $ \Delta$ " on the constructed supermanifold and
discuss
 theirs connection with the  geometrical objects on underlying manifold.

 In the  {\it Appendix 1} for the general case we briefly mention
  the method of Hamiltonian reduction in the terms convenient
 for our purposes.

In the {\it Appendix 2}  we recall the connection
 between the supermanifolds and the linear bundles to the
   extent necessary for our purposes. In this Appendix we
 suggest a natural lifting in the general case of the
 reduction procedure from the manifolds to their corresponding
  supermanifolds with odd symplectic structures.

  For  rigorous definitions and conventions in supermathematics used here
 we refer too [1].

  The preliminary results of this article were  published in [17].

       \setcounter{equation}0

		\section{ Examples of K\"ahlerian
       	                Supermanifolds and Hamiltonian  Reduction  }

        We mostly consider symplectic structures (odd
    or even one) as the part of  corresponding  K\"ahlerian structures.
    In the same way as in the bosonic case [18]   complex
    supermanifold is provided by even (odd)
    K\"ahlerian structure if symplectic
    structure is defined by
    real closed nondegenerated  even (odd) two-form $\Omega^{\kappa}$
    which   in  local complex coordinates $z^A$, is given by the
    following expression
 \begin{equation}
     \Omega^{\kappa}=i(-1)^{p(A)(p(B)+\kappa+1)}g^\kappa_{A {\bar B}}
           dz^A \wedge d{\bar z}^B,
\end{equation}
      where
$$  g^\kappa_{A {\bar B}}  =
     (-1)^{(p(A)+\kappa+1)(p(B)+\kappa+1)+\kappa +1}
         \overline {g^\kappa _{B {\bar A}}},\quad p(g^\kappa _{A\bar B})=p_A
+p_B+\kappa$$
Here and further index $\kappa =0(1)$ denote even(odd) case.

Then there exists a local real even (odd) function $K^\kappa(z,{\bar z})$
    (K\"ahlerian potential), such that
\begin{equation}
	        g^\kappa_{A {\bar B}}  =
              \frac{\partial ^L}{\partial z^A}
		\frac{\partial ^R}{\partial {\bar z}^B}
			       K^\kappa  (z,{\bar z})
\end{equation}
     (As well as in usual case [18] the potential $K$ is
   defined with precision define up to arbitrary analytic and antianalytic
   functions.)

    To even (odd) form $\Omega^{\kappa}$ there corresponds
    the even (odd) Poisson bracket
   \begin{equation}
		     \{ f,g\}_\kappa
                        =
		     i\left(
	     \frac{\partial ^R f}{\partial \bar z^A}
			   g^{{\bar A}B}_\kappa
	        \frac{\partial  ^L g}{\partial z^B}
		           -
            (-1)^{(p(A)+\kappa)(p(B)+\kappa)}
      		 \frac{\partial  ^R f}{\partial z^A}
		     g^{{\bar A}B}_\kappa
		 \frac{\partial ^L g }{\partial \bar z^B}
		       \right),
\end{equation}
 where
$$g^{{\bar A}B}_\kappa g_{B{\bar C}}^\kappa=\delta^{\bar A}_{\bar C}
\;\;,\;\;\;\; \overline{g^{{\bar A}B}_\kappa}
 = (-1)^{(p(A)+\kappa)(p(B)+\kappa)}g^{{\bar B}A}_\kappa .$$
 Its  satisfied to conditions of reality and "antisimmetricity"
       \begin{equation}
 \overline{\{ f, g\}_\kappa }=\{\bar f ,\bar g \}_\kappa ,\;\;\;\{ f, g
\}_\kappa = -
(-1)^{(p(f)+\kappa)(p(g)+\kappa)}\{ g, f \}_\kappa ,
\end{equation}
and Jacobi identities :
\begin{equation}
( -1)^{(p(f)+\kappa)(p(h)+\kappa)}\{ f,\{ g, h \}_{\kappa}\}_\kappa +{\rm
{(cicl. perm.)}} = 0
\end{equation}
On the complex superspace $\DC ^{N+1,N+1}$ with complex
   coordinates $z =(z^n,\eta ^n ),$  $n=0,1,...,N$
  canonical symplectic structure
       $$
      \Omega^0 = i(dz^n \wedge {\bar dz}^n -i d\eta^n \wedge d{\bar \eta}^ n)
$$
 with corresponding even Poisson bracket
\begin{equation}
   \{ f,g\}_0 = i\left(\frac{\partial f}{\partial z^n}
                         \frac{\partial g}{\partial {\bar z}^n}    -
                        \frac{\partial f}{\partial {\bar z}^n}
                        \frac{\partial g}{\partial z^n}\right)   +
	      \frac{\partial ^R f}{\partial \eta^n}
	      \frac{\partial  ^L g}{\partial {\bar \eta}^n} +
	     \frac{\partial ^R f}{\partial {\bar \eta}^n}
	      \frac{\partial  ^L g }{\partial  \eta^n}
\end{equation}
    defines  even K\"ahlerian structure,
and  canonical
   odd symplectic structure
$$\Omega^1 = dz^n \wedge d{\bar \eta}^n + d{\bar z}^n \wedge d\eta^n$$
     with corresronding odd bracket
   \begin{equation}
   \{ f,g\}_1 =\frac{\partial f}{\partial z^n}
                     \frac{\partial^L g}{\partial {\bar \eta}^n}    +
                     \frac{\partial f}{\partial {\bar z}^n}
                      \frac{\partial^L g}{\partial \eta^n}   -
  \frac{\partial ^R f}{\partial {\bar \eta}^n}
	\frac{\partial g}{\partial z^n} -
	      \frac{\partial ^R f}{\partial \eta^n}
	     \frac{\partial g}{\partial {\bar z}^n}
\end{equation}
defines  odd K\"ahlerian structure.
   One can obtaines more nontrivial examples by Hamiltonian  reduction.

   It is well known that for the harmonic oscillator in
   $(N+1)_{\DC}$--dimensional phase space  using the energy integral
   for decreasing by one the complex degrees of freedom we  go to
   $N$-dimensional complex projective space and
   K\"ahlerian metric corresponding to reduced symplectic structure
   on it coincides with canonical one  [19].
     The straightforward generalization of this procedure on
   supercase gives us the following example.

Let
 \begin{equation}
   	    H= z^n {\bar z}^n -i \eta^n {\bar \eta}^n
\end{equation}
    be the Hamiltonian of the superoscillator in the complex phase
   superspace $\DC^{N+1,N+1}$  with even Poisson bracket (2.6).
     $ H$ defines Hamiltonian action of group $U(1)$  on   $\DC^{N+1,N+1}$
      via motion equations
\begin{equation}
        {\dot f}=\{ H,f \}_0\quad,\quad z \rightarrow e^{it}z
\end{equation}
  As well as in the ordinary case  the $(N.N+1)$ -
   dimensional  complex projective
   superspace $\DC P(N.N+1)$ (the manifold of $(1.0)$ - dimension
    complex subspaces in the $\DC ^{(N+1.N+1)}$ ) is obtained as the
   factorization of the
    $(2N+1.2N+2)_{\DR}$ -dimensional level supersurface
\begin{equation}
                         H=h
\end{equation}
    by Hamiltonian action (2.9) of the group $U(1)$.
      One can choose  as the lo\-cal co\-or\-di\-nates
   of the su\-per\-ma\-ni\-fold  $\DC P(N,N+1)$
    in the map $z^m \neq 0$  the functions
     $w^A_{(m)}=(w^a_{(m)}, \eta ^k _{(m)}),
        a \neq m $,
where
\begin{equation}
		      w^a_{(m)}=\frac{z^a}{z^m}\quad,\quad
	     \theta^k _{(m)}=\frac{\eta ^k}{z^m}
\end{equation}
    restricted on the supersurface (2.10).
      The transition functions for these coordinates
     from the map $z^n\neq 0$ to the map
     $z^m\neq 0$ are
\begin{equation}
	    w^a_{(m)} =\frac{w^a_{(n)}}{w^m _{(n)}},  \;\;
	   \quad \theta^k _{(m)} =
          \frac{\theta^k _{(n)}}{w^m _{(n)}}, \quad {\rm where}\quad w^m_{(n)}=
     (w^a_{(n)}, w^n_{(n)}=1).
\end{equation}
    These coordinates are invariant under $U(1)$ group action:
                            $$
                   \{ w^a_{(m)} , H \}_0=
             \{ \theta^k _{(m)} , H \}_0= 0.
                          $$
        So the
    inherited Poisson bracket on $\DC P(N,N+1)$ is naturally
         defined by the relation
	      		    $$
		       \{f,g\}_0^{\rm red}=
		       \{f,g\}_0\mid_{H=h},
			                $$
    where $f,g$ are functions depending on the coordinates
    $w^A_{(m)}, {\bar w^A_{(m)}}$
    (see for details Appendix 1 or [19]).

  The calculations give us
			 \begin{eqnarray}
 \{ w^A_{(m)},w^{ B}_{(m)}\}_0^{\rm red}& =&
   \{{\bar w}^{ A}_{(m)},{\bar w}^{ B}_{(m)}\}_0^{\rm red} = 0,\nonumber \\
\{w^A_{(m)},{\bar w}^{ B}_{(m)}\}_0^{\rm red}&=&
 (-1)^{p_A p_B +1}\{{\bar w}^B_{(m)},{w}^{A }_{(m)}\}_0^{\rm red} = \\
 &= &(i)^{p_A p_B +1} \frac{1+(-i)^{p_C} w^C_{(m)} {\bar w}^{ C}_{(m)}}{h}
	     (\delta^{AB}+(-i)^{p_A p_B} w^A_{(m)} {\bar w}^{B}_{(m)}) .\nonumber
\end{eqnarray}
 From  (2.12) one obtain that the coordinates
     $w^A_{(m)}$ provide $\DC P(N.N+1)$ by complex structure
    and to Poisson bracket (2.13) correspond   K\"ahlerian
   structure with potential:
			  $$
	  K_{(m)}= h\log (1+(-i)^{p_C} w^C_{(m)} {\bar w}^{ C}_{(m)}) .
			  $$
 \vspace{0.7cm}

  Let us consider now  the reduction of the  odd Poisson bracket (2.7)
   on the $\DC^{N+1,N+1}$
    by Hamiltonian action of $U(1)$ group.
    It is easy to check that it defined by an odd Hamiltonian
   \begin{equation}
			Q_2=i(z^k {\bar \eta}^{ k} -
	    {\bar z}^{ k} \eta^{ k}),
\end{equation}
  ( which is supercharge of previous one), because it is easy to check that
  for arbitrary function $f$:
		      $$
      {\dot f}=\{H,f\}_0=\{Q_2,f\}_1
		     $$
    where $\{\quad,\quad\}_1$ is odd Poisson bracket (2.7).
   Performingng the reduction as above we obtain the supermanifold
   $M_{\DR}^{2N+1.2N+1}$ of real dimension $(2N+1,2N+1)$ which evidently
   can not has (even) complex structure.
   We define an odd  symplectic structure on it  similary to even case :
 the $U(1)$ -invariant functions
   $(w^a, \theta^a, {\bar w^a}, {\bar \theta^a}, H_0, Q_1)$ where
   $w^a, \theta^a,$ are defined by (2.11) ,
\begin{eqnarray}
		    H_0 &=& z^k {\bar z}^{ k},    \\
    		   Q_1 &=& z^k {\bar \eta}^{ k} +
		    {\bar z}^{ k} \eta^{ k}  ,
\end{eqnarray}
    restricted on the level supersurface
$$Q_2=q_2$$
can be seen as local
   coordinates of $M_{\DR}^{2N+1.2N+1}$. In these coordinates the odd Poisson
bracket is
    defined by following basic relations
$$ \{w^a_{(m)},{\bar \theta }^b_{(m)}\}_1^{\rm red}=
  \frac{i(1+ w^c_{(m)} {\bar w}^c_{(m)})}{ H_0}\delta^{ab} ,\quad
 \{\theta^a_{(m)},H_0 \}_1^{\rm red}=-w^a_{(m)} ,  \quad
\{Q_1,H_0 \}_1^{\rm red}=H_0 .$$
         We see that the same transformations group $U(1)$  of the comp\-lex
   su\-per\-space \break $\DC^{(N+1,N+1)}$  which  Hamiltonian action  in
  both cases is defined by (2.9)  reduces this
 superspace to rather different  symplectic supermanifolds.

In the following section by Hamiltonian reduction
    we construct a complex supermanifold  which can be considered
  as a reduction of both of them and which have naturally
   defined even and odd K\"ahlerian structures.

  \setcounter{equation}0
\section{ Su\-per\-ge\-ne\-ra\-li\-za\-tion of $\DC P(N)$ with Even and Odd
	   K\"ahlerian Structures}
In this section we do Hamiltonian reduction of initial
    $\DC^{N+1,N+1}$  with canonical even structure (2.6) by one
   generalization of  $U(1)$ and  the reduction of
    $\DC^{N+1,N+1}$ with canonical odd structure (2.7) by another
    generalization of $U(1)$. The complex supermanifolds obtained in both
   cases appear to be the same (up to diffeomorfism) and can be considered as
   "intersection" of supermanifolds considered above. This supermanifold
   provided by even and odd K\"ahlerian structures
   turns to be associated to the tangent  bundle of complex
   projective space $\DC P(N)$.

\subsection{Reduction by Even Bracket}

   Now let us consider at first the reduction
  of even structure (2.6) on the superspace  $\DC^{N+1,N+1}$
      by Hamiltonian $H$  and its supercharges
  $Q_1$  and $Q_2$ (which defined by (2.8), (2.14), (2.16)).
 They form the superalgebra
\begin{equation}
  \{Q_r,Q_s\}_0 =2 \delta_{rs}H,\quad \{Q_r ,H \}_0 =\{H ,H\}_0 =0
       ,\quad r,s=1,2.
\end{equation}
       This superalgebra defines the Hamiltonian action of $(1.2)$--
 dimensional group of transformations of the  $\DC^{N+1,N+1}$ .
  To every even element $\tilde H =\alpha H + \beta Q_1 +\gamma Q_2$ (where
$\alpha$
 is even
and $\beta$ and $\gamma$ is odd constants) of this
  superalgebra corresponds one-parametric transformation
  $z \rightarrow {\tilde z}(t,z)$ via motions equations
    ${\dot z}=\{\tilde H,z\}_0$.
   The  group of these transformations is the supergeneralization of the
  $U(1)$ group transformations (2.9). We denote it by $U^s(1)$.
      Lets  define in
      $\DC^{(N+1,N+1)}$ the level supersurface $M_{h,q_{1},q_{2}}$
         by equations
  \begin{equation}
	 H=h, \;\;\;\;   Q_1=q_1, \;\;\;\;  Q_2=q_2.
\end{equation}
  Reduced phase superspace is  the factorization of $M_{h,q_{1},q_{2}}$ by  the
  action of $U(1)$  subgroup of  $U^s(1)$, because transformations
  cor\-res\-pon\-ding to  $Q_1$ and to $Q_2$ do not preserve (3.2).
    For pulling down Poisson bracket (2.6) on it
    we have to choose  convenient local coordinates which are
    $U^s(1)$ -invariant functions on  $\DC^{(N+1,N+1)}$   restricted on
$M_{h,q_{1},q_{2}}$
    (see for details Appendix 1)
       These coordinates  are following
\begin{eqnarray}
            \sigma^a_{(m)}&=&-i\{w^a_{(m)},Q_+\}=
             \theta^a_{(m)}-\theta^m_{(m)} w^a_{(m)},   \\
  x^a_{(m)}&=& w^a_{(m)}+i\frac{Q_-}{H}\sigma^a_{(m)},
\end{eqnarray}
         where $w^a_{(m)}, \theta^a_{(m)}, \theta^m_{(m)}$ are defined by
(2.11)
and
$$ Q_{\pm} =\frac{Q_1 \pm iQ_2}{2}.$$
   These coordinates provide reduced superspace
by complex structure (see Subsection 3.3).

    If $f$ and $g$ are $U^s(1)$--invariant functions
   then $\{f,g\}$ is  $U^s(1)$--invariant function
   too, so from (3.1), (3.3) using Jacoby identity (2.5) one can obtain
     that their Poisson brackets  depend only on $x^a$,  ${\bar x^a}$,
   $ \sigma^a$  $ {\bar \sigma^a}$, and $H$ .
    The inherited Poisson bracket as well as in previous section
    is defined by the relation
	      		    $$
		       \{f,g\}_0^{\rm red}=
		       \{f,g\}_0\mid_{H=h,Q_{1,2}=q_{1,2}},
			                                 $$
      where  $f,g$ are $U^s(1)$ -invariant functions, $\{\quad,\quad\}_0$
    is the canonical even bracket (2.6) on $\DC^{N+1,N+1}$.
   Substituing (3.3), (3.4) in this relation and taking
  into account (2.13), (3.1), (3.2) and $U^s(1)$--invariance
   one obtain by  straightforward calculations
\begin{eqnarray}
\{x^A,x^B\}_0^{\rm red}&=&\{{\bar x}^A,{\bar x}^B\}_0^{\rm red}= 0 ,\quad {\rm
where}\quad x^A =(x^a ,\sigma^a) \nonumber \\
	      \{x^a,{\bar x^b}\}^{\rm red}_0&=&
i\frac{A}{h}(\delta^{ab}+x^a{\bar x}^b) -
       \frac{\sigma^a{\bar \sigma}^b}{h},   \nonumber \\
\{ x^a,\bar\sigma^b \}_0^{\rm red}&=&
   i\frac{A}{h}\left( x^a{\bar \sigma}^b +
     \mu (\delta^{ab}+x^a{\bar x}^b )\right)   \\
 \{\sigma^a,{\bar \sigma^b}\}^{\rm red}_0 &=&
  \frac{A}{h}\left( (1+i\mu{\bar \mu})\delta^{ab}+x^a{\bar x}^b +
 i(\sigma^a +\mu x^a)({\bar \sigma}^b+{\bar \mu}{\bar x}^b \right),\nonumber
   \end{eqnarray}
(other relations are obtaned from (3.5) taking into account (2.4))
 where
$$	 A= 1+x^a{\bar x}^a -i\sigma^a {\bar \sigma}^a +
 \frac{i\sigma^a {\bar x^a} {\bar \sigma}^b x^b}{1+x^c{\bar x}^c }, \;\;\;\;
    \mu= \frac{{\bar x}^a \sigma^a}{1+x^b{\bar x}^b } . $$

   One can show that to odd structure (3.5) corresponds
   K\"ahlerian structure with potential
\begin{equation}
   K= h\log\left(1+x^a \bar x^a - i\sigma^a\bar\sigma^a+\frac{i\sigma^a{\bar
x^a}
                    \bar \sigma^b x^b}{1+x^c \bar x^c} \right)
\end{equation}

\subsection{Reduction by Odd Bracket}
     In the same way we consider the reduction of the  $\DC^{(N+1.N+1)}$
   with odd structure (2.7) by another supergeneralization
   $U^{\tilde s}(1)$ of the group $U(1)$ generated by $Q_2$ and
   $H_0$ (defined by (2.14), (2.15))(as it was mentioned above
   $Q_2$ defines $U(1)$ group action (2.9) in terms of
  odd bracket). This group is abelian :
                         $$
            \{H_0,Q_2\}_1= \{H_0,H_0\}_1=\{Q_2,Q_2\}_1=0
                         $$
  so reduced phase superspace have real
  dimension $(2N.2N)$. The functions
  $w^a_{(m)}, \sigma^a_{(m)}$, defined
  by (2.11) and (3.3) commute with $Q_2$ and $H_0$
 so their restriction on levels
  supermanifold
$$Q_2=q_2,\quad H_0=h_0$$
 are the   appropriate local coordinates for pulling down
    odd Poisson bracket on a reduced superspace.
    The inherited odd Poisson bracket is defined in the same way as
   (3.5):
 $$\{f,g\}_1^{\rm red}=
		       \{f,g\}_{1}\mid_{H_0=h_0,Q_2=q_2},
                          $$
      where  $f,g$ are $U^{\tilde s}(1)$--invariant functions,
  $\{\quad,\quad\}_1$  is the canonical
   odd bracket (2.7) on $\DC^{N+1,N+1}$ .
      The calculations give
\begin{eqnarray}
\{w^A,w^B\}^{\rm red}_1&=&\{{\bar w}^A,{\bar w}^B\}^{\rm red}_1=0 , \quad{\rm
where} \quad w^A=(w^a,\sigma^a)\nonumber\\
\{w^a,{\bar w^b}\}^{\rm red}_1 &=& 0,  \nonumber \\
\{w^a,{\bar \sigma}^b\}_1^{\rm red} &=&
  \frac{1+w^c{\bar w}^c}{h_0}(\delta^{ab}+w^a{\bar w}^b) , \\
 \{\sigma^a,{\bar \sigma^b}\}_1^{\rm red}&=&
 \frac{1+w^c{\bar w}^c}{h_0}(\sigma^a{\bar w}^b-w^a{\bar \sigma}^b)
  +\nonumber \\
&+& \left(\frac{\sigma^c{\bar w}^c-w^c{\bar \sigma}^c}{h_0}+
    iq_2(1+w^c{\bar w}^c)\right)(\delta^{ab}+w^a{\bar w}^b) \nonumber.
\end{eqnarray}
 (other relations are obtaned from (3.7) taking into account (2.4)).
    Corresponding odd K\"ahlerian structure  (the local coordinates
$(w^a_{(m)},\sigma^a_{(m)})$
 provide reduced superspace
by complex structure (see Subsection 3.3)).
is given by potential
 \begin{equation}
			    K_1=
		          h_ 0\frac
     {i(w^a {\bar \sigma^a}-{\bar w^a}\sigma^a)}{1+w^b{\bar w^b}}
			     +
		  q_2 \log (1+w^a{\bar w^a}) .
 \end{equation}

\subsection{ Investigation of the Global Properties}

	      We obtain two reduced superspaces one
   with coordinates  $x^a, \sigma^a$ and even K\"ahlerian structure
with potential (3.6),
  another  with coordinates  $w^a, \sigma^a$
and odd K\"ahlerian structure with potential (3.8) .
    Now we show that they coincide up to diffeomorphism and
  clarify their global structure.
  It is not useless for these purposes to investigate
  the transitions functions from map to map for
  coordinates  $w^a_{(m)}, \sigma^a_{(m)}$
   and  $x^a_{(m)}, \sigma^a_{(m)}$.

    The coordinates $\sigma^a_{(m)}$  transform like differentiatiales
   of the $w^a_{(m)}$ according their definition (3.3).
\begin{eqnarray}
 w^a_{(n)}\rightarrow  w^a _{(m)}
       & =&\frac{ w^a _{(n)}}{w^m _{(n)}} , \\
     \sigma^a _{(n)}\rightarrow  \sigma^a _{(m)}
 &=&\frac{\sigma^a_{(n)}w^m _{(n)} - w^a _{(n)}\sigma^m_{(n)}}
       {(w^m _{(n)})^2} , \quad k=0,..., N. \nonumber
\end{eqnarray}
where $(w^n _{(n)} =1 ,\;
 \sigma^n_{(n)}=0 )$.

 From (3.4) and (3.9)
   it is easy to see that the coordinates $(x^a_{(m)}, \sigma^a_{(m)})$
    transform like $(w^a_{(m)}, \sigma^a_{(m)})$:
\begin{eqnarray}
 x^a_{(n)}\rightarrow x^a_{(m)}&=&\frac{x^a _{(n)}}{ x^m _{(n)}},\\
\sigma^a_{(n)}\rightarrow \sigma^a_{(m)}
    & =& \frac{\sigma^a_{(n)}x^m _{(n)} - x^a _{(n)}\sigma^m_{(n)}}
      {(x^m _{(n)})^2} \quad (x^n_{(n)}=1,\sigma^n_{(n)}=0)  \nonumber
\end{eqnarray}
As seen, this supermanifolds have global complex structures.

    It allows us to consider these two reduced superspaces
    as the same  because one can identify
    $(w^a_{(m)}, \sigma^a_{(m)})$ with $(x^a_{(m)}, \sigma^a_{(m)})$.
   The correspondence
   $(x^a_{(m)},\sigma^a_{(m)})\rightarrow (w^a_{(m)},\sigma^a_{(m)})$
  preserving under the  transformations (3.9), (3.10) sets up isomorphism
  from the functions defining on the reduced superspace with even
   structure (3.5) on  the functions defining on the reduced superspace with
  odd structure (3.7).
The obtained phase superspace we denote by $ S\DC P(N) $. \\
 Now let us summarize our results .
The phase superspace  $\DC P(N.N+1)$ which was constructed
   in the Section 2  as the reduction of $\DC^{N+1.N+1}$
    with even canonic structure by the Hamiltonian of
   superoscillator (without using its supercharges)
   and  now  constructed  $S\DC P(N)$ have the same
   underlying manifold - $N$-dimensional complex
  projective space  $\DC P(N)$ . The K\"ahlerian
  structure which corresponded to (2.13) on the $\DC P(N.N+1)$ as well as the
  even K\"ahlerian structure with (3.5) for $S\DC P(N)$ pull down
  to the standard K\"ahlerian structure of underlying
  complex projective space.   $S\DC P(N)$ can be
    considered as the further reduction of  $\DC P(N.N+1)$ by
    the supercharges.
	 In contrary to $\DC P(N)$  it have naturally
     defined odd K\"ahlerian structure with potential (3.8) and can
be considered  as further reduction of $M_{\DR}^{2N+1.2N+1}$ by $H_0$ too
$$  \DC^{(N+1.N+1)} { \buildrel H\over \longrightarrow}\DC P(N.N+1)
  {\buildrel  {Q_1,Q_2}\over \longrightarrow} S\DC P(N)
                \quad{\rm ({even \quad reduction})}  $$
$$  \DC^{(N+1.N+1)}{\buildrel  {Q_2}\over
          \longrightarrow}M^{2N+1.2N+1}_{\DR}
   {\buildrel {H_0}\over \longrightarrow}S\DC P(N)
 \quad{\rm {(odd \quad reduction})}.      $$
 Moreover from the equations (3.9), (3.10) it is easy to see that
   $S\DC P(N)$ with local coordinates
   $x^a_{(m)}, \sigma^a_{(m)}$  is associated
   to the $T\DC P(N)$ - tangent bundle
  of the underlying manifold $\DC P(N)$ because the even coordinates from
   map to map transform through themself only
   and odd coordinates transform as
    differentials of even ones [1] (see also Appendix 2).\\

 From this point of view it becomes natural the following
 property of the odd symplectic structure (3.7).
 One can show that in the coordinates
		     $$
     	  {\tilde \sigma}_a =  g_{a\bar b}\bar{\sigma^b}   ,
		  		     $$
   where  $g_{ab}$ is the  K\"ahlerian metric of the underlyind projective
   space  ,
    the odd symplectic structure turns out to be canonical one if $Q_2 =0$
(for general case, if $Q_2 \neq 0 $ see Appendix 2).
                          $$
  {\tilde \Omega}^1 = dw^a \wedge d{\tilde \sigma}_a + {d\bar w^a}\wedge
        {d\bar {\tilde \sigma}_a}$$
   Indeed in the coordinates  $(w^a ,\sigma_a)$ $S\DC P(N)$ is associated
   to $T^* \DC P(N)$ - cotangent bundle of $\DC P(N)$, which have naturally
  defined canonical symplectic structure [19].\\

    It has been mentioned in Introduction that these constructions
  have general meaning.
   Indeed for every K\"ahlerian manifold $M$ with local
  complex coordinates $w^a$   one can consider
  the complex supermanifold
  ${SM}$ ($dim_{\DC} SM = (dim_{\DC}M. dim_{\DC}M)$)
  with   local coordinates  $w^a, \sigma^a$
   which is associated to $TM$. Then the local functions
\begin{eqnarray}
K_0(w,{\bar w},\sigma,{\bar \sigma})&=& K(w,{\bar w})+
  F(ig_{a{\bar b}}(w,{\bar w})\sigma^a{\bar \sigma}^b),  \quad p( K_0 )=0\\
  K_1(w,{\bar w},\sigma,{\bar \sigma})&=&
    \epsilon\frac{\partial K(w,{\bar w})}{\partial w^a}\sigma^a+
  {\bar \epsilon}\frac{\partial K(w,{\bar w})}
   {\partial {\bar w^a}}{\bar \sigma^a}+\alpha K(w,\bar w),\quad p( K_1 )=1
   \end{eqnarray}
 ( where $K(w,\bar w)$ is the K\"ahlerian potential
of $M$ , $g_{a\bar b}$- corresponding Riemannian metric,
  $F(r)$-arbitrary scalar function such that $F'(0) \neq 0$,
 $\epsilon$ is even complex constant an $\alpha$ is real odd one)
 can be considered as
  the potentials which correctly define global even and odd K\"ahlerian
structures  on $SM$ [16].

  In the case $M=\DC P(N)$ we obtain immediately the structures
constructed above putting in (3.11), (3.12) $ K(w,\bar w)=\log (1+w^a \bar w^a
),
 F(r)=\log (1-r), \epsilon=i, \alpha=q_2$.
\setcounter{equation}0

\section{Operator $\Delta$ and bi-Hamiltonian Mechanics }
	  Now we want to discuss the properties of some supergeometrical
  constructions which can be  defined in natural way on the
 supermanifolds provided by even and odd symplectic structures
  studying them on the supermanifold constructed above.

  The supermanifolds which are associated in some coordinates
 to tangent bundle (see Appendix 2)
  can be considered as "gauge fixing" objects for the studying
 the supergeometrical constructions  which in this case have to reduce
 to the well-known geometrical objects. So this constructions
 can be considered as the generalization on
 supercase of the corresponding geometrical objects.

  From this point of view it is interesting to look at
 the explicit expressions for  the "operator $\Delta$" and the bi-Hamiltonian
mechanics on the $S\DC P(N)$ provided by odd and even brackets (3.6), (3.9)
(similar expressions for the
 supermanifolds provided by two K\"ahlerian structures with potentials
 (3.11), (3.12) see in [16]).

   \subsection{ Operator $ \Delta$ on $S\DC P(N)$ }
On the su\-per\-ma\-ni\-fold ${\cal M}^{m.m}$
  with co\-or\-di\-na\-tes $z^A=(x^1,\ldots,x^{m},\theta^1,\ldots,\theta^m)$
  which is provided by odd symplectic structure with Poisson bracket
$\{\quad,\quad\}_1$ and
 the volume form $dv=\rho(x, \theta)d^{m}x d^{m}\theta$
  one can invariantly define the odd differential operator of the
 second order, so called "operator $ \Delta$" which is invariant
 under the transformations preserving the symplectic structure
  and the volume form [2, 11]. Its action on the function $f(x,\theta)$
  is the divergence  of the Hamiltonian
  vector field ${\bf D}_{f} =\{f,z^A \}_1\frac{\partial^L}{\partial z^A}$
  with the volume form $dv$:
\begin{equation}
\Delta f =div ^{\rho} {\bf D}_{f} \equiv \frac{{\cal L}_{{\bf D}_f} dv}{dv},
\end{equation}
where  ${\cal L}_{{\bf D}f}$ --- Lie derivative
along ${{\bf D}_f}$ [1].
In coordinate form
\begin{equation}
           \Delta f=\frac{1}{\rho}(-1)^{p(A)}
     \frac{\partial^L}{\partial z^A}\left(\rho\{z^A,f\}_1\right)
\end{equation}
   The "operator $ \Delta$" have no analogs with even symplectic structures
--- the oddness of the Poisson bracket  $\{\quad,\quad\}_1$
  which force that operator (4.2) to have dependence of second derivatives.

If the density  $\rho=1$ and $\{\quad,\quad\}_1$ has the canonical
  form (1.2)
  then $\Delta$  is in
 the canonical form
\begin{equation}
\Delta ^{\rm can}  f =2\frac{\partial ^2 f}{\partial x^i \partial \theta ^i}
,
\end{equation}
 which is well-known from BV-formalism [2-4].

It is easy to obtain from (4.1) using Jacobi identities, Leibnitz
 rules and the transformation law of integral density $\rho(z)$
 that generelized operator $\Delta$ (4.2) is connected with
corresponding odd bracket by the same expressions
as  canonical  operator $\Delta^{\rm can}$ (4.3) connected with canonical odd
bracket (1.2) [3, 4]:
\begin{eqnarray}
\Delta \{f,g \}_1& =&\{f,\Delta g \}_1 +(-1)^{p(g)+1}\{\Delta f ,g \}_1
\nonumber \\
(-1)^{p(g)}\{f,g \}_1 &=&\frac{1}{2}\left ( \Delta(fg) - f\Delta g
-(-1)^{p(g)}(\Delta f)g  \right )\nonumber \\
 \Delta'f &=& \Delta f +\{\log{{\cal J}} ,f \}_1 ,\nonumber
\end{eqnarray}
where ${\cal J}$-Jacobian of canonical transformation of odd bracket,
$\Delta'$-
operator $\Delta$ in new coordinates.
However the nilpotency condition
\begin{equation}
	       \Delta^2=0
\end{equation}
are violated for arbitrary $\rho (x,\theta)$.

For example , if symplectic structure is canonical, (4.4) hold if
$\rho (x,\theta)$ satisfy to the equation
$$\Delta \rho =0$$
which is  master equation of BV-formalism for the action $S=\log \rho$ .
Then $\Delta$ corresponding to operator of BRST transformation [2-4].
  It is interesting to study the connection between the condition (4.4)
  and the possibility to reduce (4.2) to (4.3) by the suitable
   transformation of the coordinates.

  If  the supermanifold ${\cal M}$ provided by even symplectic
structure $\Omega^0$ also  here one can
 put into (4.2) the density $\rho$ , which is invariant under canonical
transformations of $\Omega^0 $  [19, 20]:
\begin{equation}
	   \rho(z)=\sqrt{{\rm Ber} \Omega_{AB}}.
\end{equation}
Let ${\cal M} = S\DC P(N)$ provided by odd Poisson bracket (3.7) ( with $q_2
=0$)
and even one (3.5).The invariant (under canonical transformations
of (3.5)) density $\rho$ on it has the  form
\begin{equation}
 \rho(w, \bar w, \theta, \bar \theta) =(1-r)^2 \equiv (1-ig_{a\bar b}\theta^a
\bar{\theta^b})^2
\end{equation}
 where
\begin{equation}
g_{a\bar b}=\frac{1}{1+w^c \bar w^c}\left (\delta_{a\bar b} -
\frac{\bar w^a w^b}{1+w^c \bar w^c} \right)
\end{equation}
--K\"ahlerian metric of $\DC P(N)$ ($r$ corresponds
to  cohomologies   on $\DC P(N)$) .
The operator $\Delta$ on $S\DC P(N)$ with this density takes the folowing form
    \begin{equation}
	    \Delta f = \frac{1}{\rho}
		 \left(\nabla^a \frac{\partial^L}{\partial\theta^a} +
		 {\overline {\nabla^a}}\frac{\partial^L}{\partial{\bar \theta}^a}\right)(\rho
f),
\end{equation}
      where
$$ \nabla_a=\frac{\partial}{\partial w^a} -
        \Gamma^c_{ab}\theta^b\frac{\partial^L}{\partial\theta^c},\quad
           {\overline {\nabla^a}}=g^{{\bar a}b}\nabla_b ,$$
$\Gamma^c_{ab}=g^{\bar d c}g_{a \bar d ,b} \equiv
-\frac{\bar w^a \delta^c _b +\bar w^b \delta^c_a}{1+w^d \bar w^d}$ --
the Christoffel symbols of the
  K\"ahlerian metric (4.7) on $\DC P(N)$.
 Nilpotency condition (4.4) is satisfied obviously .
 The operator (4.8) corresponds to the operator
of covariant divergency $\delta = \ast d \ast$ on $\DC P(N)$.

 Since ${\cal M}=SM$ with K\"ahlerian
potentials (3.11), (3.12) ($\epsilon =i, \alpha = 0$) operator $\Delta$ is
also defined by the expression (4.8)  [16], where $\Gamma^c_{ab}$--
the Christoffel symbols of the  K\"ahlerian metric on underlieng manifold $M$,
	       \begin{equation}
	     \rho= \frac{{\rm det}(\delta^a_b+iF^{\prime}(r)
	      R^a_{bc{\bar d}}\theta^c\theta^d)}
	     {F^{\prime}(r)^{N-1}(F^{\prime}(r) +F^{\prime\prime}(r)r)}
\end{equation}
where $ R^a_{bc\bar d}=(\Gamma^a_{bc})_{,\bar d}$ is the curvature tensor on
$M$,
$r=ig_{a\bar b}$.
(We see that $\rho$ depends on Chern classes of the underlying
 K\"ahlerian manifold.It is interesting to compare (4.9) with the general
formulas
 for characteristic classes on the supermanifolds [20].)
To operator $\Delta$ on $SM$ is corresponds the covariant divergence
 on the un\-der\-lying K\"ah\-le\-ri\-an ma\-ni\-fold  $M$.
 \subsection{ Bi-Hamiltonian Mechanics on $S\DC P(N)$}
  Here we deliver explicit formulae for the even vector fields
  preserving even and odd
 Poisson brackets (bi-Hamiltonian mechanics) (3.5), (3.7)  on ${S\DC P(N)}$.
  In other words
 we have to find the pairs   of the functions $(H,Q)$ ($p(H)=0$,
 $p(Q)=1$) on $S\DC P(N)$ such that for arbitraty function  $f:$
   \begin{equation}
		   \{H,f\}_0 = \{Q,f\}_1,
\end{equation}
   where $\{\quad,\quad\}_0$ ($\{\quad,\quad\}_1$) defines by (3.5), (3.7).
To every pair $(H,Q)$ the
  solution of (4.10) corresponds vector field
$${\bf D}_{H,Q} =\{H,z^A \}_0\frac{\partial^L}{\partial z^A}=\{Q,z^A
\}_1\frac{\partial^L}{\partial z^A} $$
  These  fields form a finite-dimensional Lie algebra [13] and they are
 defined by Killing vectors of the underlying manifold $M$ [16].
   The solutions of the (4.10)
 is following:
 \begin{eqnarray}
 H&=&H_0 -
 \frac{i}{1-r}\frac{\partial^2 H_0}{\partial w^a \partial {\bar w}^b}
	      \theta^a{\bar \theta}^b, \nonumber\\
     Q&=&i\left(\frac{\partial H_0}{\partial w^a}\theta^a -
 \frac{\partial H_0}{\partial {\bar w}^a}{\bar \theta}^a\right),\nonumber
\end{eqnarray}
    where
$$   H_0=\frac{h_{a{\bar b}}w^a {\bar w}^b - {\rm tr} h + h_a w^a +
   {\overline {h_a w^a}}} {1+w^c{\bar w}^c},
$$
    $h_{a{\bar b}}$ are arbitrary Hermitian matrices and
  $h_a$ -- arbitrary complex numbers.
Corresponding vector field
\begin{equation}
{\bf D}_{H, Q} = V^a (w)\frac{\partial}{\partial w^a}+ V^a_c(w)\theta ^c
\frac{\partial}{\partial \theta ^a} ,
\end{equation}
where
$$   V^{a}(w)=ig^{\bar b a}\frac{\partial H_0(w,\bar w)}{\partial \bar w^b}$$
is the Killing vector of $\DC P(N)$.
Since ${\bf D}_{H, Q}$ defined by (4.11) is holomorphic
and Hamiltonian for the both
brackets, it is the Killing vector for both K\"ahlerian structures.

Bi-Hamiltonian mechanics on superanifold $SM$ with symplectic structures,
defining by (3.11), (3.12) have a similar form (4.11), where $V^a$ is
Killing vector of underlying K\"ahlerian manifold $M$ [16].

\section{Acknowledgments}
We would like to thank  A.V. Karabegov for useful
discussions , R.L. Mkrtchian and V.I. Ogievetsky for interest to this work.
\appendix
\section{1. On Procedure of Hamiltonian Reduction }
      In this Appendix we retell the main algebraic notions
  of Hamitonian reduction mechanism using  language which is
  maximally adapted for our purposes and can be evidently generalized
  on supercase.(The detailed considerations see in [19]).
   Let $M$ be symplectic space with symplectic structure
    $\Omega$ and $\Gamma(M)$  be an algebra of functions on $M$ .
    Poisson bracket  $\{\quad,\quad\}$ corresponding to $\Omega$
  defined by the following relation
   $\{f,g\} = \Omega({\bf D}_f,{\bf D}_g)$  where ${\bf D}_f$ is the vector
  field corresponding to $f$ via the equation
			$$
              \Omega({\bf D}_f, {\bf V})=df({\bf V})
					       		       $$
    for  arbitrary vector field ${\bf V}$.

     Let $C$  be an subalgebra in   $\Gamma(M)$   which is closed under
      $\{\quad,\quad\}$ .
     The algebra of functions $\Gamma(M)$
        has two algebraic operations -- usual multiplicative
  structure and Lie algebra multiplication provided by Poisson bracket
     $\{\quad,\quad\}$. Further if it is not pointed
  we suppose the first operation only.)

   Let the functions $F_1,\ldots,F_k$  be generators of $C$. In this case
     $\{F_i,F_j\}=c_{ij}^k F_k$ where $c_{ij}^k$ are constants,
     the functions $\{F_i\}$ generate Hamiltonian
   action of the group $G$ (corresponding to Lie algebra
  with structure constants $c_{ij}^k$) on the M .
    To every function $F_i$ corresponds
   $G$ group infinitesimal transformation via the vector field ${\bf D}_{F_i}.$
     and ${\bf D}_{\{F_i,F_j\}}=[{\bf D}_{F_i},{\bf D}_{F_i}]$.

     To subalgebra $C$ corresponds the reduction procedure from
     $M$ to symplectic manifold $M^{red}$.

    Let  $M_p$ be the level manifold in $M$ defined by
                            $$F_i=p_i  $$
   and $G_p$ -- its isotropy group: $G_p=\{g\in G: gp_ig^{-1}=p_i\}$.
  Then $M^{\rm red}=M_p /G_{p_i}$  and $\Omega$ is pulling down
   on the $\Omega^{\rm red}$ on $M^{\rm red}$ defining its symplectic
structure.

   For supercase it is more convenient to describe $M^{\rm red}$
  and  $\Omega^{\rm red}$ correspondingly in the terms
  of $\Gamma^{\rm red}$ -- algebra of the functions on it  and
  $\{\quad,\quad\}^{\rm red}$ -- Poisson bracket corresponding to
   $\Omega^{\rm red}$. (The generators of  $\Gamma^{\rm red}$ are the
coordinates
  of $M^{\rm red}$.)
 Let $B(M)$ be an subalgebra of the functions which is
   ``orthogonal" to subalgebra $C$ by Poisson bracket $\{\quad,\quad\}$.
  $$	B=\{\Gamma\ni f: \{f,g\}=0 \quad\forall g\in C\}  $$
  in other words $f\in B$ iff $\{f,F_i\}=0$.

 Because of Jacoby identity $B$ is Lie algebra too:
 $$ f,g\in B\Rightarrow \{f,g\}\in B, $$
  so $B$ is the subalgera of $G$ --- invariant functions of $\Gamma$.

   To level manifold $M_p$ corresponds the ideal
   ${\cal J}$ in the algebra $\Gamma$ generating by the functions
   $F_i - p_i$
 $${\cal J}=\{\Gamma \ni f: f=\sum \alpha_i(F_i - p_i)
     \quad{\rm where}\quad \alpha_i\in \Gamma\}  $$
   $B\cap{\cal J}$ is the ideal in B too, so on can consider subalgebra
		   $$ \Gamma^{\rm red}=B/B\cap {\cal J}. $$
  It is the algebra of functions on reducted space $M^{\rm red}$.
  The Poisson bracket $\{\quad,\quad\}^{\rm red}$ on $ \Gamma^{\rm red}$
   is defined in the following
  way. For any $[f],[g]\in B$, where $[f]$ is the equivalence
  class of the function $f\in B$ in $ \Gamma^{\rm red}$
  $$  \{[f],[g]\}^{\rm red} = [\{f,g\}]. $$
      To check the correctness of this definition we note
   that if $f,g\in B$ then $\{f,g\}\in B$ too . If
   $f\rightarrow {\tilde f}=f+h$ where
   $h= \sum {\alpha_i(F_i - p_i)}\in B\cap {\cal J}$ then
  $\{h,g\}\in B$ and
 $$          \{h,g\}=\{\sum {\alpha_i (F_i - p_i)}, g\}=
   \sum \{\alpha_i, g\}(F_i -p_i) \in {\cal J}
                            $$
  because $\{F_i-p_i,g\}=0$ . So $\{{\tilde f},g\}-\{f,g\}
   \in B\cap {\cal J}$.

  The reduction procedure leads to the fact that if
   dynamical system on $M$
  is described by Hamiltonian $H$ which is $G$ - invariant ($H\in B$)
   and at $t=0$ the conditions $F_i = p_i$ hold then

  i) these conditions preserve in a time,

  ii) $[f^t]=[f]^t$,

   where $h^t$ we denote the evolution of the function $h$ in the time
   $t$ via motion equations ${\dot h}=\{H,h\}$
  $[{\dot h}]=\{[H],[h]\}^{\rm red}$ .

   As example we retell in these terms the reduction procedure
  performed in the Subsection 3.1.

  We consider as $C$ the algebra of functions on the $\DC^{(N+1.N+1)}$
  which explicitly depend on the functions $H,Q_1,Q_2$  playing the role
  of generators $F_i$. The "orthogonal" subalgebra  $B$
   of $U^s(1)$ - invariant functions  is
  the algebra of functions explicitly depending on $x^a$,  ${\bar x^a}$,
   $ \sigma^a$  $ {\bar \sigma^a}$,   and $H$ functions.
   The functions $f(H-h)+g(Q_1-q_1)+r(Q_2-q_2)$ where $f,g,r$ are arbitrary
   functions consist the ideal ${\cal J}$ . $B\cap {\cal J}$ -
  the  $U^s(1)$ - invariant part of this ideal consists on the functions
   depending only on $H$.
   So the generators of the algebra  $\Gamma^{red}=B/B\cap {\cal J}$
   are
  $[x^a]$,  $[{\bar x^a}]$, $[ \sigma^a]$  $ [{\bar \sigma^a}]$
 and the functions (coordinates)
  $x^a$,  ${\bar x^a}$, $ \sigma^a$  $ {\bar \sigma^a}$
  are their corresponding representatives.
\appendix
\section{2. Supermanifolds and Linear Bundles}

      In this Appendix we briefly mention the connection
 between supermanifolds and corresponding linear bundles
 to the extent necessary for our  purposes.
 (See in details in [1].)

    Let $TM$ be the tangent bundle to the manifold $M$.
  $x^a_{(m)}$ are the local coordinates on the $M$ in $m$-th map
  and the  $(x^a_{(m)}, v^a_{(m)})$ are the corresponding
  local coordinates on $TM$ ( $v^a_{(m)}$ are coordinates of tangent
 space in the basic $\frac{\partial}{\partial x^a_{(m)}}$).
 From map to map
			$$
	x^a_{(k)} \rightarrow x^a_{(m)} =x^a_{(m)} (x^a_{(k)}),\quad
       v^a_{(k)} \rightarrow v^a_{(m)} =
  \frac{\partial x^a_{(m)}}{\partial x^b_{(k)}}v^b_{(k)}.
                                                         \eqno (A2.1) $$
 Considering for every map the superalgebra generating by
  $(x^a_{(m)}, \theta^a_{(m)})$ where $x^a_{(m)}$ are even and
  $\theta^a_{(m)}$ are odd,  transforming from map
 to map like $(x^a_{(m)}, v^a_{(m)})$ in the (A2.1)
  ($v\leftrightarrow \theta$) we go to
 supermanifold ${\cal M}$ which is associated to $TM$ in the
  coordinates $(x^a_{(m)}, \theta^a_{(m)})$
  For the coordinates $(x^a_{(m)}, \theta^a_{(m)})$ on the
  ${\cal M}$  the more general class  of transformations is admittable :
                           $$
x^a\rightarrow {\tilde x}^a(x^a,\theta^a) \quad
	\theta^a\rightarrow {\tilde \theta}^a(x^a,\theta^a)    $$
   which do not correspond to (A2.1).
 In particularly if $\theta^a \to {\tilde \theta}_a= g_{ab}\theta^b,$
  where $g_{ab}$ is some Riemanian metric on $M$ then the supermanifold
  ${\cal M}$ in the coordinates $(x^a,{\tilde \theta}_a )$ is
  associated to the cotangent bundle $T^*M$ of $M$.

 On the supermanifolds which can be associated in some
  coordinates to tangent or cotangent bundle the
  superstructures evidently are reduced to the standard geometrical objects.

     For example on the supermanifold ${\cal M}$ considered here
 the canonical odd (Buttin) bracket $\{\quad,\quad\}_1$ (defined by
 basic relations   $\{x^i , {\tilde \theta}_j\}_1=\delta^i_j )$
  is corresponding
 to the Schouten bracket $[\quad,\quad]$ of the polyvector fields on $M$:
 To polyvector field ${\bf T}= T^{j_1,\ldots,j_k}(x)$ on $M$ corresponds the
  function  $\rho {\bf T} = T(x,\theta)=
  T^{j_1,\ldots,j_k}(x)\theta_{j_1}\ldots\theta_{j_k}$
   on the ${\cal M}$ such that
 			   $$
 		\{\rho{\bf T},\rho{\bf U}\}_1= \rho[{\bf T},{\bf U}] .
 			  $$
   Similarly  operator $D=\theta^a\frac{\partial}{\partial x^a}$
  on the ${\cal M}$ corresponds to the exterior differentiation operator
  on $T^*M$ and operator $\Delta $ to the divergence [1 ].

   On one hand these type supermanifolds can be served as
 the good tests for studying superstructures on other hand
  we can use them as condensed language for constructed
 the geometrical structures in superterms. We deliver
 one example which is strightly connected with the
 considerations in the Subsection 3.2.

 The reduction procedure performed in Section 3 was indeed
  the prolongation  of the
  $M{\buildrel {\rm {reduction}}\over \rightarrow}M^{\rm red}$
  to the
  $TM{\buildrel {\rm {reduction}}\over \rightarrow}TM^{\rm red}$
  in the case $M = \DC ^{N+1}$, $M^{\rm red} = \DC P(N).$

Now for the odd structure
  reduction we show that in the general case.
  Let $M$ be the symplectic manifold with symplectic structure
 defined by Poisson bracket $\{\quad,\quad\}$ and the functions
   $I_r.$  generate Hamiltonian action
  of the Lie group $G$ on it:
			   $$ \{I_r,I_s\}=c^t_{rs}I_m   $$
    where $c^t_{rs}$ are the structure constants of
 the Lie algebra ${\cal G}$ of $G$.
  Let $M^{\rm red}$ be the manifold obtained by reduction:
  $M^{\rm red} = M_p /G_p$ where $M_p =\{x\in M: I_r(x)=p_r\}$ is the level
manifold
  and $G_p$ - is its isotropy group.

   Let $x^i$ be the local coordinates
  on $M$ and $y^a$ -
  the local ones on $M^{\rm red}$ in which the reduction was performed :
  $\{y^a(x), I_r(x)\}=0$. Then
               $$
     w^{ab} =\{y^a, y^b\}^{\rm red} = \{y^a(x), y^b(x)\}\mid_{I_r (x)=p_r}
							     \eqno(A2.2)
                $$
  defines the reduced Poisson bracket (and symplectic structure)
on $M^{\rm red}$.

  If ${\cal M}$ is supermanifold associated to $T^{*}M$ in the  local
  coordinates
   $(x^i,\theta_i)$ and Poisson bracket $\{\quad,\quad\}_1$  defines
  the odd canonical structure on it  then  it is easy to see that
  the functions
$$Q_r=\{I_r(x),x^i\}\theta_i=\{I_r,F\},\quad {\rm where}\quad
 F=\frac{1}{2}\{x^i,x^j\}\theta_i\theta_j $$
    define the same Hamiltonian action
 of the group $G$ on the $M$ in the terms of odd bracket:
 for arbitrary function $f(x)$ on $M$
 $$ \{f(x),I_r(x)\}=\{f(x),Q_r(x,\theta )\}_1.     $$
  Moreover the functions $(Q_r, I_r)$
    define the Hamiltonian action of the supergeneralization
 of the group $G$ on the ${\cal M}$ in the terms of odd bracket:
			$$
   \{Q_r, Q_s\}_1=c^t_{rs}Q_t, \quad  \{Q_r, I_s\}_1=c^t_{rs}I_t,
   \quad    \{I_r, I_s\}_1=0 .                               $$

One can show that the functions $y^A =(y^a, \eta^a =\{y^a(x), x^i\}\theta_i
=\{y^a (x), F(x, \theta)\}\_1 )$ play
the role of local coordinates on $( dim M^{\rm red}. dim M^{\rm
red})$--di\-men\-si\-onal
reduced supermanifold ${\cal M}^{\rm red}$ :
  $$  \{y^A(x,\theta ),I_r \}_1 =\{y^A (x,\theta ),Q_r \}_1 =0, $$
and in this coordinates ${\cal M^{\rm red}}$ associated to $TM^{\rm red}$.
The functions $y^A =(y^a, \eta^a)$ one can used for reduction of odd bracket
$\{\quad ,\quad \}_1$ on ${\cal M}^{\rm red}$:
$$  \{y^a,y^b\}^{\rm red}_1 =0,\quad
  \{y^a,\eta^b\}^{\rm red}_1 =w^{ab},$$
 $$  \{\eta^a,\eta^b\}^{\rm red}_1 = \frac{\partial  w^{ab}}{\partial
y^c}\eta^c
   +\frac{\partial  w^{ab}}{\partial p^r} q^r   $$
   where $w^{ab}(y,p)$ is given by (A2.2) and $I_r=p_r,\;\;
   Q_r=q_r$ define the level supermanifold in  ${\cal M}$.

  One can construct local coordinates  $(y^a,{\tilde \eta}_a)$ such that
      in these coordinates   ${\cal M}^{\rm red}$
 is associated to $ T^{*}M^{\rm red}$ :
			 $$
{\tilde \eta}_a = w_{ab}\eta^b -\frac{\partial A_a}{\partial p^k} q^k,$$
   where
  $$w_{bc}w^{ca}=\delta^a_b \;\;\; {\rm and} \;\;\;
 \frac{\partial A_a}{\partial y^b}-
 \frac{\partial A_b}{\partial y^a} =w_{ab}.$$
In this coordinates reduced symplectic structure coincides with canonical one :
 $$   \{y^a,y^b\}^{\rm red}_1=0,\quad  \{\tilde \eta^a ,\tilde \eta^b\}^{\rm
red}_1 =0 ,\\
     \{y^a,{\tilde \eta}^b\}^{\rm red}_1=\delta^a_b\quad .		 $$

\end{document}